\documentstyle[12pt]{article}
\def\Re{{\cal R \mskip-4mu \lower.1ex \hbox{\it e}\,}}
\def\Im{{\cal I \mskip-5mu \lower.1ex \hbox{\it m}\,}}
\def\ie{{\it i.e.}}
\def\eg{{\it e.g.}}

\def\etal{{\it et al.}}

\def\sub#1{_{\lower.25ex\hbox{$\scriptstyle#1$}}}
\def\sul#1{_{\kern-.1em#1}}
\def\sll#1{_{\kern-.2em#1}}
\def\sbl#1{_{\kern-.1em\lower.25ex\hbox{$\scriptstyle#1$}}}
\def\ssb#1{_{\lower.25ex\hbox{$\scriptscriptstyle#1$}}}
\def\sbb#1{_{\lower.4ex\hbox{$\scriptstyle#1$}}}

\def\gev{\,{\rm GeV}}

\def\to{\rightarrow}
\def\mh{\ifmmode m\sbl H \else $m\sbl H$\fi}
\def\mch{\ifmmode m_{H^\pm} \else $m_{H^\pm}$\fi}
\def\mt{\ifmmode m_t\else $m_t$\fi}
\def\mc{\ifmmode m_c\else $m_c$\fi}
\def\mz{\ifmmode M_Z\else $M_Z$\fi}
\def\mw{\ifmmode M_W\else $M_W$\fi}
\def\mws{\ifmmode M_W^2 \else $M_W^2$\fi}
\def\mhs{\ifmmode m_H^2 \else $m_H^2$\fi}
\def\mzs{\ifmmode M_Z^2 \else $M_Z^2$\fi}
\def\mts{\ifmmode m_t^2 \else $m_t^2$\fi}
\def\mcs{\ifmmode m_c^2 \else $m_c^2$\fi}
\def\mchs{\ifmmode m_{H^\pm}^2 \else $m_{H^\pm}^2$\fi}
\def\ztwo{\ifmmode Z_2\else $Z_2$\fi}
\def\zone{\ifmmode Z_1\else $Z_1$\fi}
\def\mtwo{\ifmmode M_2\else $M_2$\fi}
\def\mone{\ifmmode M_1\else $M_1$\fi}
\def\tb{\ifmmode \tan\beta \else $\tan\beta$\fi}
\def\xw{\ifmmode x\sub w\else $x\sub w$\fi}
\def\ch{\ifmmode H^\pm \else $H^\pm$\fi}
\def\lum{\ifmmode {\cal L}\else ${\cal L}$\fi}
\def\inpb{\ifmmode {\rm pb}^{-1}\else ${\rm pb}^{-1}$\fi}
\def\infb{\ifmmode {\rm fb}^{-1}\else ${\rm fb}^{-1}$\fi}
\def\epem{\ifmmode e^+e^-\else $e^+e^-$\fi}
\def\ppb{\ifmmode \bar pp\else $\bar pp$\fi}

\def\bsg{\ifmmode b\rightarrow s\gamma \else $b\rightarrow s\gamma$\fi}

\newskip\zatskip \zatskip=0pt plus0pt minus0pt
\def\matth{\mathsurround=0pt}

\def\atversim#1#2{\lower0.7ex\vbox{\baselineskip\zatskip\lineskip\zatskip
  \lineskiplimit 0pt\ialign{$\matth#1\hfil##\hfil$\crcr#2\crcr\sim\crcr}}}

\renewcommand{\thefootnote}{\fnsymbol{footnote}}

\begin{document} \begin{titlepage}
\setcounter{page}{1}
\thispagestyle{empty}
\rightline{\vbox{\halign{&#\hfil\cr
&SLAC-PUB-6427\cr
&January 1994\cr
&T/E\cr}}}
\vspace{0.8in}
\begin{center}

{\Large\bf
Constraints From $b \to s\gamma$ on the Left-Right Symmetric Model}
\footnote{Work supported by the Department of
Energy, contract DE-AC03-76SF00515.}
\medskip

\normalsize THOMAS G. RIZZO
\\ \smallskip
{\it {Stanford Linear Accelerator Center\\Stanford University,
Stanford, CA 94309}}\\

\end{center}

\begin{abstract}

Recent results from the CLEO Collaboration on both inclusive and exclusive
radiative $B$ decays are used to constrain the parameter space of two versions
of the Left-Right Symmetric Model. In the first scenario, when the left- and
right-handed Cabibbo-Kobayashi-Maskawa mixing matrices are equal,
$V_L=V_R$, the radiative $B$ decay data is shown to lead to strong bounds
on the $W_L-W_R$ mixing angle that are quite insensitive to either the
top quark or $W_R$ mass. The second scenario examined is that of Gronau
and Wakaizumi wherein $b$-quark decays proceed only via right-handed currents
and $V_L$ and $V_R$ are quite distinct. For this model, the combined
constraints from Tevatron $W_R$ searches, the $B$ lifetime, and radiative
$B$ decays lead to a very highly restricted allowed range for the $W_L-W_R$
mixing angle.

\end{abstract}

\vskip0.30in
\begin{center}

Submitted to Physical Review {\bf D}

\end{center}


\renewcommand{\thefootnote}{\arabic{footnote}} \end{titlepage}


While the Standard Model(SM) of strong and electroweak interactions is in very
good agreement with all existing experimental data{\cite {leppho}}, there are
many reasons to believe that new physics(NP) must exist not far above the scale
currently being probed at the SLC, LEP, and Tevatron colliders. Although we
do not know what form this NP might take there are a vast number of
proposals in the literature. The best that we can do in the `pre-discovery'
era is to use existing data to restrict the properties of this NP and
to continue searching. While colliders provide us with the capability to
directly produce signatures of NP, a complementary approach is to hunt for NP
indirectly through high precision measurements and the observation of rare
processes. An excellent working example of such a process has been provided
us by the CLEO Collaboration
{\cite {cleo}} which has recently observed the exclusive decay
$B \to K^* \gamma$ with a branching fraction of $(4.5 \pm 1.5 \pm 0.9)
\times 10^{-5}$ and has placed an upper limit on the inclusive quark-level
process of $B(b \to s\gamma)<5.4 \times 10^{-4}$ at $95\%$ CL. Using a
conservative estimate for the ratio of exclusive to inclusive decay rates
{\cite {soni}}, the observation of the exclusive process implies the
{\it {lower}} bound $B(b \to s\gamma)>0.60 \times 10^{-4}$ at the $95\%$ CL.
These values are, of course, consistent with SM expectations{\cite {rev}} but
can used to restrict various forms of NP, as has been done in the recent
literature{\cite {big}}. It is important to note that both the upper {\it {as
well as}} the lower bounds can be used to constrain NP since any model leading
to an extremely suppressed rate for this process is already excluded by the
CLEO data. Not all of the analyses{\cite {big}} have taken advantage of this
additional constraint.

One scenario of NP which has been popular in the literature for many years and
has had many manifestations is the Left-Right Symmetric Model(LRM)
{\cite {lrm}} based on the extended electroweak gauge group $SU(2)_L \times
SU(2)_R \times U(1)$. Amongst other things, this model predicts the existence
of a
heavy, right-handed, charged gauge boson, $W_R^{\pm}$, which can in principle
mix through an angle $\phi$ with the more conventional $W_L^{\pm}$ present in
the SM to form the mass eigenstates $W_{1,2}$. Data from, \eg, polarized
$\mu$ decay{\cite {mu}} (in the case of light right-handed neutrinos) and
universality requirements{\cite {sir}} tell us
that the size of this mixing must be reasonably small (less than, say,
$|\phi|=0.05$, or so) but whose
exact magnitude depends on the detailed assumptions we make about the other
features of the model{\cite {lang}}. As we will see below, the exchange
of $W_R^{\pm}$ within a
penguin diagram, in analogy with the SM $W$ exchange, can lead to significant
deviations from SM predictions for the $b \to s\gamma$ branching fraction
which is quite sensitive to the both the sign and magnitude of $\phi$.

In order to numerically determine the branching fraction for $b \to s\gamma$
within the
LRM there are several sets of parameters whose values we need to address:
$(i)$ the mass of the $W_R$ itself; $(ii)$ the ratio of the right-handed
to left-handed $SU(2)$ gauge
group coupling constants, \ie, $\kappa=g_R/g_L$; $(iii)$ the mixing
angle $\phi$; and $(iv)$ the numerical values for the elements of the mixing
matrix $V_R$. For purposes of our discussion below we will treat $\phi$ as a
free parameter and use the data on the $b \to s\gamma$ decay itself to
constrain $\phi$ as a function of the other
degrees of freedom. {\it If} we assume $V_L=V_R$ then there are several strong
constraints on the $W_R$ mass arising from both collider searches{\cite {wr}}
as well as the $K_L-K_S$ mass difference{\cite {lang,soni2}} and it is likely
that $M_{W_R}>1.6\kappa$ TeV. Although this possibility is both simple
and attractive, realistic and
phenomenologically viable models can be constructed wherein $V_R$ and $V_L$
are quite unrelated as in the scenario of Gronau and Wakaizumi(GW)
{\cite {gron}}
that we will discuss in more detail below. In such models, at least some of
the conventional constraints on the $W_R$ mass can be evaded. However, all
such bounds are also dependent on the value of $\kappa$ and,
within the context of grand unified theories, we generally find
that $\kappa \leq 1${\cite {desh}}. One might naively expect in a more
general context that this ratio or couplings differ from unity by no more
than a factor of two or so.

The approach we follow in performing our calculations has already been
discussed in our earlier work{\cite {us}} and we will refer the interested
reader to those papers for calculational details. An outline of this approach
is as follows. To obtain the \bsg\
branching fraction, the inclusive $b \to s\gamma$ rate is scaled to that of
the semileptonic decay $b\to X\ell\nu$.  This removes major
uncertainties in the calculation associated with ($i$) an overall factor of
$m_b^5$ which appears in both expressions and ($ii$) the various right- and
left-handed
Cabibbo-Kobayashi-Maskawa (CKM) factors.  We then make use of the data
on the semileptonic branching fraction{\cite {drell}}, which is given by
$B(b \to X \ell \nu)=0.108$, to rescale our result.
The semileptonic rate is calculated including both phase space(due to the
large value of $m_c/m_b$) and QCD corrections{\cite {cab}}
with $m_b=5\gev$ and $m_c=1.5\gev$.   The calculation of
$\Gamma(\bsg)$ employs the next-to-leading log evolution equations
for the coefficients of the $b\to s$ transition
operators in the effective Hamiltonian due to
Misiak{\cite {misiak}}, the gluon bremsstrahlung corrections of Ali and
Greub{\cite {ali}}, the leading corrections from heavy quark effective theory
(HQET){\cite {falk}}, a running $\alpha_{QED}$ evaluated at the b-quark mass
scale, and 3-loop evolution of the running $\alpha_s$ matched to the value
obtained at the $Z$ scale via a global analysis{\cite {leppho}} of all data.
As we will see, the bounds we obtain on the parameters of the LRM are not
very sensitive to the remaining uncertainties{\cite {uncer}} in the
calculation of the $b \to s \gamma$ branching fraction arising from higher
order QCD corrections. In what follows we limit our attention to the
contributions of the charged gauge bosons to the $b \to s \gamma$ decay rate.
In principle, there are potentially other significant contributions in the LRM
owing to the extended nature of the symmetry breaking sector, \ie, there
can be significant contribution from charged Higgs exchange as well as from
flavor-changing neutral Higgs exchange; we will ignore both these
possibilities in the analysis below.

To complete
the calculation we use the one-loop matching conditions for the
various operators{\cite {misiak}} in a form that includes contributions from
both the SM and new LRM operators, \ie, for every `left-handed' operator
present in the SM, the existence of light-right symmetry dictates the
existence of the corresponding `right-handed' one. The two sets of operators
do not mix under QCD evolution and can thus be treated independently. The
$b \to s \gamma$ branching fraction can then be expressed as, using
$\alpha^{-1}_{QED}(m_b)=132.7$,
\begin{equation}
B(b\to s\gamma)= {{6\alpha_{QED}(m_b)}\over {\pi(1+Q)}} B(b\to c\ell\nu)
{{|C^{eff}_{7L}|^2+|C^{eff}_{7R}|^2}\over{(L^2_\ell+R^2_\ell)
[(L^2_h+R^2_h)f+2L_hR_hg]}}F \,,
\end{equation}
where $Q(f,g)$ is the QCD(phase space) correction to the semileptonic decay
$b \to c \ell \nu$. While $Q$ as a function of $m_c/m_b$ is given
in{\cite {cab}}, the explicit forms for $f$ and $g$ are given, \eg, in
{\cite {lepton}:
\begin{eqnarray}
f & = & (1-y^4)(1-8y^2+y^4)-24y^4\ln y \,, \\
g & = & -2y[(1-y^2)(1+10y^2+y^4)+12y^2(1+y^2)\ln y] \,, \nonumber
\end{eqnarray}
where $y=m_c/m_b$. For $y=0.3$, we obtain $f \simeq 0.520$,
$g \simeq -0.236$, and $Q \simeq 2.50({2\over 3\pi}\alpha_s(m_b))$. The
factor $F$ denotes the relatively small corrections from HQET and
gluon bremsstrahlung mentioned
above and are both of order a few percent. Defining $t_{\phi}=tan \phi$ and
$r=(M_{W_1}/M_{W_2})^2$ (using $M_{W_1} \simeq 80.21$ GeV in numerical
calculations), we obtain for a general LRM
\begin{eqnarray}
(L_\ell^2+R_\ell^2)(L^2_h+R^2_h) & = & |V^L_{cb}|^2[(1+rt_\phi^2)^2
+\kappa^2t_\phi^2(1-r)^2] \nonumber \\
    &  & + |V^R_{cb}|^2[\kappa^2t_\phi^2(1-r)^2+\kappa^4(r+t_\phi^2)^2] \,, \\
2L_hR_h(L^2_\ell+R^2_\ell) & = & 2\kappa t_\phi(1-r)\Re (V^L_{cb}V^R_{cb})
[(1+rt_\phi^2)+\kappa^2(r+t_\phi^2)] \,. \nonumber
\end{eqnarray}
(Note that we implicitly assume that the mass of the right-handed neutrino is
sufficiently low as to allow its participation in the B decay process and lets
us neglect corrections of order $m_{\nu_R}^2/m_b^2$.)
$C_{7L,R}^{eff}$ are defined via the low-energy effective Hamiltonian
\begin{equation}
{\cal H}_{eff}=-{G_Fem_b\over 4\sqrt 2\pi^2}\bar s\sigma_{\mu\nu}
\Bigl( C^{eff}_{7L}P_R+C^{eff}_{7R}P_L \Bigr) bF_{\mu\nu} \,,
\end{equation}
where $P_{L,R}=(1\pm \gamma_5)/2$, and whose numerical values are obtained
from the operators evaluated at the weak scale($\simeq M_{W_1}$) via
a renormalization group analysis. This analysis is, of course, quite similar
to that performed for in the SM case except for the additional operators
that are present and have non-zero co-efficients at the weak scale. Of course,
in either the SM or LRM, only a few of these weak scale operator
co-efficients are
non-zero to one loop order. Assuming that the top(t)-quark contribution
dominates the penguin diagrams (as will be the case in the scenarios we
examine below), we obtain in the usual notation
\begin{eqnarray}
C_{2L}(M_{W_1}) & = & (1+rt_\phi^2)(V_{cb}V^*_{cs})_L \,, \nonumber \\
C_{2R}(M_{W_1}) & = & \kappa^2(r+t_\phi^2)(V_{cb}V^*_{cs})_R \,, \nonumber \\
C_{10L}(M_{W_1}) & = & \kappa t_\phi(1-r){m_c\over m_b}
(V_{cb}^LV^{*R}_{cs}) \,, \nonumber \\
C_{10R}(M_{W_1}) & = & C_{10L}(M_{W_1}) (L\leftrightarrow R) \,, \\
C_{7L}(M_{W_1}) & = & (V_{tb}V^*_{ts})_L[A_1(x_1)+rt_\phi^2A_1(x_2)]+
{m_t\over m_b} rt_\phi(V_{tb}^RV_{ts}^{*L})[A_2(x_1)-rA_2(x_2)] \,, \nonumber
\\
C_{7R}(M_{W_1}) & = & {m_t\over m_b}\kappa t_\phi(V_{tb}^LV_{ts}^{*R})
[A_2(x_1)-rA_2(x_2)]+\kappa^2(V_{tb}V_{ts}^*)_R[t_\phi^2A_1(x_1)+rA_1(x_2)] \,,
\nonumber
\end{eqnarray}
where $x_{1,2}=m_t^2/M_{W_{1,2}}^2$. The co-efficients of the operators
corresponding to the gluon penguin, $C_{8L,R}(M_{W_1})$, can be expressed in
a manner similar to $C_{7L,R}(M_{W_1})$ but with $A_i \to B_i$; note that
both $A_1$ and $B_1$ are the same functions found in the usual SM
calculation. Explicitly, we find
\begin{eqnarray}
A_1(x) & = & -{1\over 2}(x-1)^{-4}\Biggl[ Q_t\Biggl( {1\over 4}x^4-{3\over
2}x^3
+{3\over 4}x^2+{x\over 2}+{3\over 2}x^2\ln x\Biggr) \nonumber \\
       &  &\quad \quad \quad \quad  + \Biggl( {1\over 2}x^4+{3\over 4}x^3
-{3\over 2}x^2+{1\over 4}x-{3\over 2}x^3\ln x\Biggr)\Biggr] \,, \\
A_2(x) & = & {1\over 2}(x-1)^{-3}\Biggl[ Q_t\Biggl( -{1\over 2}x^3-{3\over
2}x+2
+3x\ln x\Biggr) \\
    &  & \quad \quad \quad \quad +\Biggl(-{1\over 2}x^3+6x^2-{15\over 2}x+
2-3x^2\ln x\Biggr)\Biggr] \,, \nonumber
\end{eqnarray}
where $Q_t=2/3$ is the top-quark electric charge and $B_{1,2}(x)$ are given
by the terms proportional to $Q_t$ in $A_{1,2}(x)$. An important feature to
note in the expressions above is the chiral enhancement, by a factor of
$m_t/m_b \sim 30$, of the terms which involve mixing between the $W_L$ and
$W_R$ gauge bosons which are proportional to a factor of $t_{\phi}=tan \phi$.
This
implies that the decay rate for $b \to s \gamma$ should be quite sensitive
to small values of $t_{\phi}$ even when the $W_2$ is quite massive.

We note in passing that the assumption of top-quark dominance of the penguin
diagrams may not always be valid in a general LRM since, in principal, the
values of the elements of both $V_L$ and $V_R$ may conspire to suppress this
contribution. This happens, however, in only a very small region of the
parameter space since $m_t/m_c>100$.

Let us first consider the situation where $V_L=V_R$; in this case
the implied lower
bound on the $b \to s \gamma$ branching fraction plays no r\^ole in
restricting the LRM parameters. If we assume that $\kappa=1$
and $M_{W_R}$ is large, we can ask for the bound on $t_{\phi}$ as a function
of $m_t$ that results from the CLEO limits; this is shown in Fig.~1
for a $W_R$ of mass 1.6 TeV and which explicitly
displays the $b \to s \gamma$ branching fraction as a function of $t_{\phi}$.
Here we see that ($i$) the constraint on the value of $t_{\phi}$ is relatively
insensitive to $m_t$ and, at $95\%$ CL, lies in the approximate range
$-0.02 < t_{\phi} <0.005$. These bounds are much more restrictive than what
one obtains from either $\mu$ decay data ($-0.056 < t_{\phi} <0.040$)
{\cite {mu}} or universality arguments ($-0.065 <t_{\phi} <0.065$)
{\cite {sir}}. ($ii$) For top masses larger than 120 GeV the $b \to s \gamma$
branching fraction($B$) is always found to be in excess of $1.4\times 10^{-4}$.
These results are found to be quite insensitive to the particular values
chosen for either the
$W_R$ mass or $\kappa$ so long as the $W_R$ is reasonably heavy.
One may wonder if in
fact we can turn this argument around in order to get a constraint on
$M_{W_R}$ itself from the CLEO data. To address this issue, we fix $m_t=160$
GeV with $\kappa=1$ and display $B$ for various values of $M_{W_R}$ as a
function of $t_{\phi}$ as shown in Fig.~2. Here we see that $B$ itself is not
very sensitive to the $W_R$ mass for fixed $m_t$ so that no limit is
obtainable from this decay mode. Lastly, for fixed $m_t=160$ GeV and $M_{W_R}=
1.6$ TeV, we can explore the sensitivity of the resulting bounds on $t_{\phi}$
as $\kappa$ is varied; this is shown in Fig.~3, for $0.6 < \kappa <2$. As might
be expected the bound strengthens with increasing values of $\kappa$, but only
weakly so for positive values of $t_{\phi}$. The strengthening of the bounds
for negative $t_{\phi}$ is much more noticeable.  It is clear from these
figures that the CLEO results provide an additional important constraint on
the LRM parameters when $V_L=V_R$ is assumed and that QCD uncertainties at the
level of $10-20\%$ will not significantly influence the results we have
obtained.

Let us now turn to the perhaps more interesting scenario of Gronau and
Wakaizumi (GW) wherein B decays proceed {\it only} via the right-handed
currents. For concreteness we take the forms of $V_L$ and $V_R$ as they appear
in the original work of GW{\cite {gron}}:
\begin{eqnarray}
V_L & = & \left( \begin{array}{ccc}
1 & \lambda & 0 \\
-\lambda & 1 & 0 \\
0 & 0 & 1
\end{array} \right) \,, \\
V_R & = & \left( \begin{array}{ccc}
c^2 & -cs & s \\
{s(1-c)\over\sqrt 2} & {c^2+s^2\over\sqrt 2} & {c\over\sqrt 2} \\
{-s(1+c)\over\sqrt 2} & -{c-s^2\over\sqrt 2} & {c\over\sqrt 2}
\end{array} \right) \,, \nonumber
\end{eqnarray}
where $\lambda (\simeq 0.22)$ is the Cabibbo angle and $s \simeq 0.09$.
In order to satisfy
B lifetime constraints, the parameters in the GW model must satisfy the
additional requirement
\begin{equation}
M_{W_R}\leq 416.2\, \kappa\left[ {|V^R_{cb}|\over\sqrt 2}\right]^{1/2} \gev
\simeq 415\, \kappa\gev \,,
\end{equation}
which arises from recent determinations of $V_{cb}$ in the SM{\cite {vcb}}.
In addition, to satisfy $\mu$ decay data, the right-handed neutrino must be
sufficiently massive ($\simeq 17$ MeV) but this has little effect on the B
decay itself. Of course, a $W_R$ satisfying the above constraint is relatively
light and should have a significant production cross section at the Tevatron
given the form of $V_R$.
In our earlier work we showed that a $W_R$ in the GW model
can satisfy {\it both} the low
energy and collider bounds provided that $\kappa \geq 1.5$ and $M_R \geq
600$ GeV{\cite {wr}} if we assume that the $W_R$ decays only into the
known SM particles as well as the right-handed neutrino.
We will respect these conditions when considering the predictions of this
model for the $b \to s \gamma$ decay, but we should remember that these
collider-based limits are
clearly softened if additional decay modes of the $W_R$ are allowed.

First, let us fix both $M_{W_R}$ and $\kappa$ in order to satisfy the above
constraints and examine the predicted value of $B$ in the GW model as a
function of $t_{\phi}$; this is shown in Figs.~4 and 5 for various values of
$m_t$. From these two figures we learn that ($i$) the allowed range of
$t_{\phi}$ is more tightly restricted in comparison to the $V_L=V_R$ case by
{\it both} the upper and lower CLEO limits. ($ii$) The bounds are quite
insensitive to $m_t$ and ($iii$) the value $t_{\phi}=0$ is almost excluded by
the CLEO lower limit. Further, we see as $M_{W_R}$ is increased (also
increasing $\kappa$ to satisfy the constraints above) we see that the curves
become steeper and this forces the allowed regions of $t_{\phi}$ to become
quite pinched and narrow. In the $M_{W_R}=600(800)$ GeV case the allowed
ranges for $t_{\phi}$ are found to be
$-0.43\times 10^{-3} <t_{\phi}< 0$ and $0.40 \times 10^{-3}<t_{\phi}<0.81\times
10^{-3}$(
$-0.32\times 10^{-3} <t_{\phi}< 0$ and $0.29 \times 10^{-3}<t_{\phi}<0.60\times
10^{-3}$). To say the least, these ranges are highly restrictive and it is
clear that a more precise determination of the value of $B$ may rule out the
model as it now stands. To show just how pinched these curves become with
increasing $M_{W_R}$, we fix $m_t=160$ GeV and let $M_{W_R}=400\kappa$ GeV
while varying $\kappa$; this is shown in Fig.~6. Clearly, as $M_{W_R}$ grows,
the allowed ranges become {\it extremely} tight and only a very fine tuning of
the parameters will allow the GW model to remain phenomenologically viable
unless other sources of new physics are introduced.

It is, of course, possible that a modified version of the GW scheme may be
realized by slightly different versions of both $V_L$ and $V_R$ and several
such scenarios exist in the literature. Hou and Wyler{\cite {hou}} have, in
fact, two distinct versions of these matrices, denoted by I and II. The
resulting predictions for the $b \to s \gamma$ branching fraction, $B$, in
both scenarios are quite similar and are shown in Figs.~7a and 7b for
$\kappa=1.5$ and $M_{W_R}=600$ GeV. (The collider bounds on the $W_R$ in
both these scenarios are essentially identical to the original GW model.)
Qualitatively, these predictions are very similar to those of the original GW
scheme. Quite recently, Hattori \etal ~have proposed another possible version
of the GW scenario{\cite {hatt}} leading to the predictions for $B$
in Fig.~8, again assuming $\kappa=1.5$ and $M_{W_R}=600$ GeV. In this
model the `no-mixing' possibility, $t_{\phi}=0$, is completely excluded by the
CLEO data, but otherwise the results are similar to that of the original GW
model. It would seem that a general result of the GW approach is to restrict
$t_{\phi}$ to very small, but most likely non-zero, values. As in
the standard GW case, the $t_{\phi}$
dependence in both the Hou and Wyler as well as the Hattori \etal ~models
becomes somewhat
stronger as the $W_R$ mass is increased to 800 GeV and $\kappa$ is set to 2.

In this paper we have examined the predictions of the Left-Right Symmetric
Model for the $b \to s \gamma$ branching fraction in the limit where only
the $W_L^{\pm}$ and $W_R^{\pm}$ gauge bosons contribute to the penguin
amplitudes.  We examined two specific versions of this model, the first,
wherein left-right symmetry is explicit and $V_L=V_R$, and the second, in
which the $b$-quark essentially decays only through right-handed currents.
This corresponds to models of the kind first constructed by Gronau and
Wakaizumi.
In the $V_L=V_R$ case, the limits we obtained on the $W_L-W_R$ mixing angle,
$\phi$, were found to be relatively independent of the top quark mass and the
assumed value of $M_{W_R}$ provided $\kappa=1$. For fixed top and $W_R$
masses, however, the sensitivity of these constraints to
variations in $\kappa$ was found to be
significant. The bounds we obtained on $\phi$ were comparable to, yet
somewhat better
than, those obtainable from $\mu$ decay data or universality arguments.
No limit on $M_{W_R}$ is obtained from these considerations
alone and only the CLEO upper bound was needed to obtain the resulting
constraints. In the GW-type scenarios, both upper and lower bounds on the
$b \to s \gamma$ branching fraction provided important input and were
folded together with additional constraints arising from
Tevatron collider searches as well as the B lifetime. Again, the
resulting limits on $\phi$ were relatively $m_t$-independent and extremely
tight, falling into two distinct regions in all the cases we examined. An
improvement in the CLEO bounds could conceivably rule out this scenario if
our assumptions remain valid, except, perhaps, for some extremely fine-tuned
cases. Additional penguin contributions in the form of, \eg, Higgs bosons,
would then be needed to rescue this approach.

Perhaps rare B decays may yet provide us with a signature for new physics
beyond the Standard Model.

\vskip.25in
\centerline{ACKNOWLEDGEMENTS}

The author would like to thank J.L.\ Hewett, E.\ Thorndike, Y.\ Rozen,
M.\ Gronau, P.\ Cho, M.\ Misiak, D.\ Cocolicchio and R.\ Springer for
discussions related to this work. The author would also like to thank the
members of the Argonne National Laboratory High Energy Theory Group for use of
their computing facilities.

Note added: After this manuscript was essentially completed we received several
preprints by various authors who have also analyzed the decay rate for
$b \to s \gamma$ in the LRM{\cite {other} for the $V_L=V_R$ case. Where
we overlap, the results we have obtained are in agreement with these other
authors.

\newpage

%
\def\MPL #1 #2 #3 {Mod.~Phys.~Lett.~{\bf#1},\ #2 (#3)}
\def\NPB #1 #2 #3 {Nucl.~Phys.~{\bf#1},\ #2 (#3)}
\def\PLB #1 #2 #3 {Phys.~Lett.~{\bf#1},\ #2 (#3)}
\def\PR #1 #2 #3 {Phys.~Rep.~{\bf#1},\ #2 (#3)}
\def\PRD #1 #2 #3 {Phys.~Rev.~{\bf#1},\ #2 (#3)}
\def\PRL #1 #2 #3 {Phys.~Rev.~Lett.~{\bf#1},\ #2 (#3)}
\def\RMP #1 #2 #3 {Rev.~Mod.~Phys.~{\bf#1},\ #2 (#3)}
\def\ZP #1 #2 #3 {Z.~Phys.~{\bf#1},\ #2 (#3)}
\def\IJMP #1 #2 #3 {Int.~J.~Mod.~Phys.~{\bf#1},\ #2 (#3)}

\newpage

%
{\bf Figure Captions}
\begin{itemize}

\item[Figure 1.]{
$b \to s\gamma$ decay mode in the LRM assuming $V_L=V_R$ as a function of the
tangent of the $W_L-W_R$ mixing angle, $t_{\phi}$. Here we assume $\kappa=1$
and a $W_R$ mass of 1.6 TeV for top-quark masses of $m_t$=120(dots), 140
(dashes), 160(dash-dots), 180(solid), or 200(square dots) GeV. The horizontal
solid line is the CLEO upper bound.}
\item[Figure 2.]{Same as Fig.~1 but with $m_t=160$ GeV held fixed and $M_{W_R}$
varied between 300(lower curve) and 1000(upper curve) GeV.}
\item[Figure 3.]{ Same as Fig.~1 with $m_t=160$ GeV and $M_{W_R}$=1.6 TeV
with $\kappa$ varying between 0.6(left-most dotted curve) and 2(inner-most
dash dotted curve).}
\item[Figure 4.]{Predicted values for the branching fraction($B$) of the
$b \to s\gamma$ decay mode in the Gronau-Wakaizumi version of the LRM as a
function of the tangent of the $W_L-W_R$ mixing angle, $t_{\phi}$. In this
figure, $\kappa=1.5$ and $M_{W_R}=600$ GeV is assumed and the
outer(inner)-most solid
line corresponds to $m_t=120(200)$ GeV and is increased in each case by steps
of 20 GeV. The dashed horizontal lines are the
CLEO upper and lower bounds.}
\item[Figure 5.]{Same as Fig.~4 but with $\kappa=2$ and $M_{W_R}=800$ GeV.}
\item[Figure 6.]{Same as Fig.~4 but for $m_t$ fixed at 160 GeV and $M_{W_R}=
400\kappa$ GeV. Here $\kappa$ is varied between 1 and 2 in steps of 0.2 with
$\kappa=1(2)$ corresponding to the outer(inner) curve.}
\item[Figure 7.]{Predicted values for $B$ in the Hou-Wyler parameterization
of $V_L$ and $V_R$ assuming $\kappa=1.5$ and $M_{W_R}=600$ GeV. The
dotted(dashed, dash-dotted, solid, square-dotted) curve corresponds to
$m_t=120(140, 160, 180, 200)$ GeV: (a)version I and (b)version II.}
\item[Figure 8.]{Same as Fig.~7, but for the parameterization of
Hattori \etal.}

\end{itemize}

\end{document}